\documentclass[12pt]{article}

\usepackage{amsmath}
\usepackage{graphicx}
\usepackage{geometry}
\usepackage{gensymb}

\begin{document}
\title{\textbf{Transmission of High-Power Electron Beams Through Small Apertures}}
\date{}

\maketitle

\author{C. Tschal\"{a}r$^1$, R. Alarcon$^4$, S. Balascuta$^4$, S.V. Benson$^2$, W. Bertozzi$^1$, J.R. Boyce$^2$, R. Cowan$^1$, D. Douglas$^2$, P. Evtushenko$^2$, P. Fisher$^1$, E. Ihloff$^1$, N. Kalantarians$^3$, A. Kelleher$^1$, R. Legg$^2$, R.G. Milner$^1$, G.R. Neil$^2$\\\ L. Ou$^1$, B. Schmookler$^1$, C. Tennant$^2$, G.P. Williams$^2$, and S. Zhang$^2$}\\\  

$^1 Laboratory for Nuclear Science, Massachusetts Institute of Technology,$

$Cambridge, MA 02139$

$^2 Free Electron Laser Group, Thomas Jefferson National Accelerator Facility,$ 

$Newport News, VA 23606$

$^3 Department of Physics, Hampton University, Hampton, VA 23668$

$^4 Department of Physics, Arizona State University, Glendale, AZ 85306$

\abstract
Tests were performed to pass a 100 MeV, 430 kWatt c.w. electron beam from the energy-recovery linac at the Jefferson Laboratory's FEL facility through a set of small apertures in a 127 mm long aluminum block. Beam transmission losses of 3 p.p.m. through a 2 mm diameter aperture were maintained during a 7 hour continuous run.

\section{Introduction}

The beam transmission test described in detail in this paper and summarized in a letter$^{[1]}$ was motivated by design studies of window-less high-density gas targets for scattering experiments with high-power electron beams. It was assumed that the beam would enter and exit the target through short, small-diameter  tubes. The target gas leaking through these tubes would be pumped away in stages to maintain vacuum in the beam pipes.
To minimize size and cost of these pumps and maximize the gas target density, the tube diameters need to be minimized. At the same time, beam losses in traversing the tubes need to be kept extremely small to minimize background.

\section{Transmission Test}

\subsection{Test setup}

The transmission tests were carried out with the 100 MeV electron beam from the energy-recovery linac at the Jefferson Laboratory's FEL facility.

At the modified F3 region of the FEL beam (see Fig. 1) between two quadrupole triplets, a remotely controllable aperture block made of aluminum containing three apertures of 2, 4, and 6 mm diameter and 127 mm length was mounted in the beam pipe (see Fig. 2).

\begin{figure}[htbp]
\centering\includegraphics[width=0.5\textwidth,height=0.25\textheight]{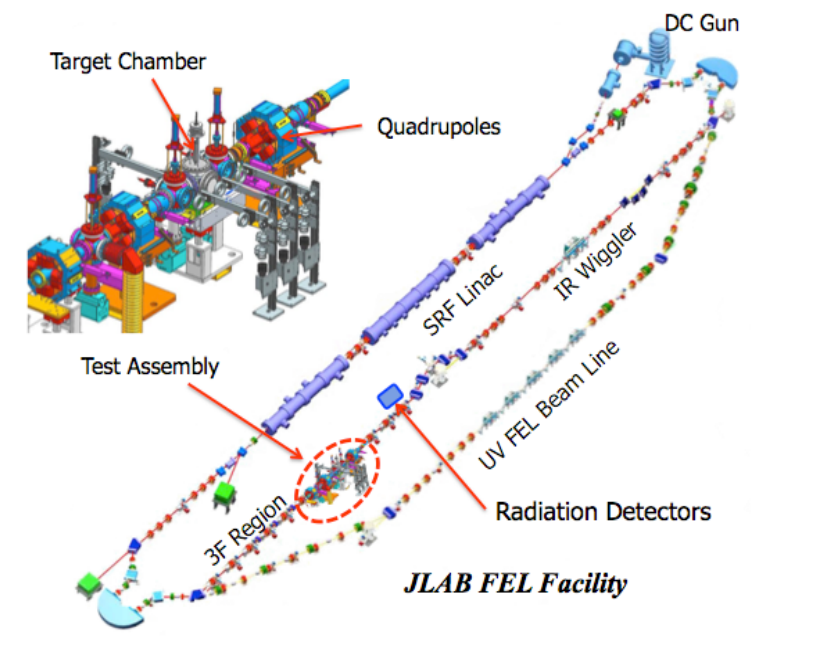}
\caption{FEL beam facility at the Jefferson Laboratory}
\label{DL_transp1}
\end{figure}

\begin{figure}[htbp]
\centering\includegraphics[width=0.5\textwidth,height=0.15\textheight]{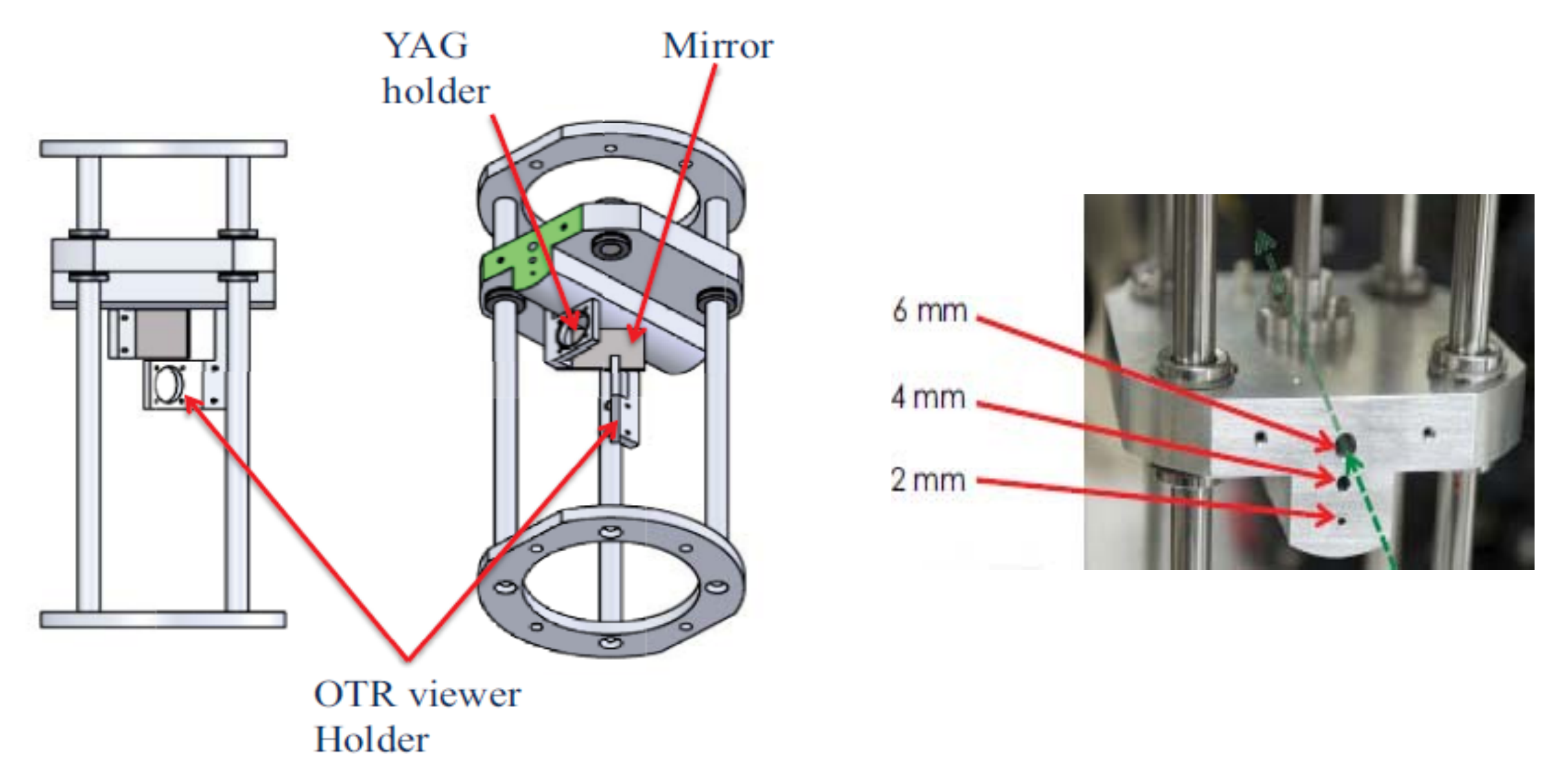}
\caption{Aperture block with 2, 4, and 6 mm apertures}
\label{DL_transp2}
\end{figure}

The block also carried a YAG crystal and an OTR crystal viewed by TV cameras to measure beam profiles and beam halo at the position of the aperture block. Any of these apertures or profile monitors could be placed on the beam axis by remote control. The temperature of the aperture block was monitored by a resistance temperature detector. The block temperature, beam current, repetition rate, and bunch charge were recorded and logged.
Neutron and photon background monitors were placed near the aperture block and around the beam lines and the linac (see Fig.3). All readings were logged$^{[2]}$.

\begin{figure}[htbp]
\centering\includegraphics[width=0.8\textwidth,height=0.2\textheight]{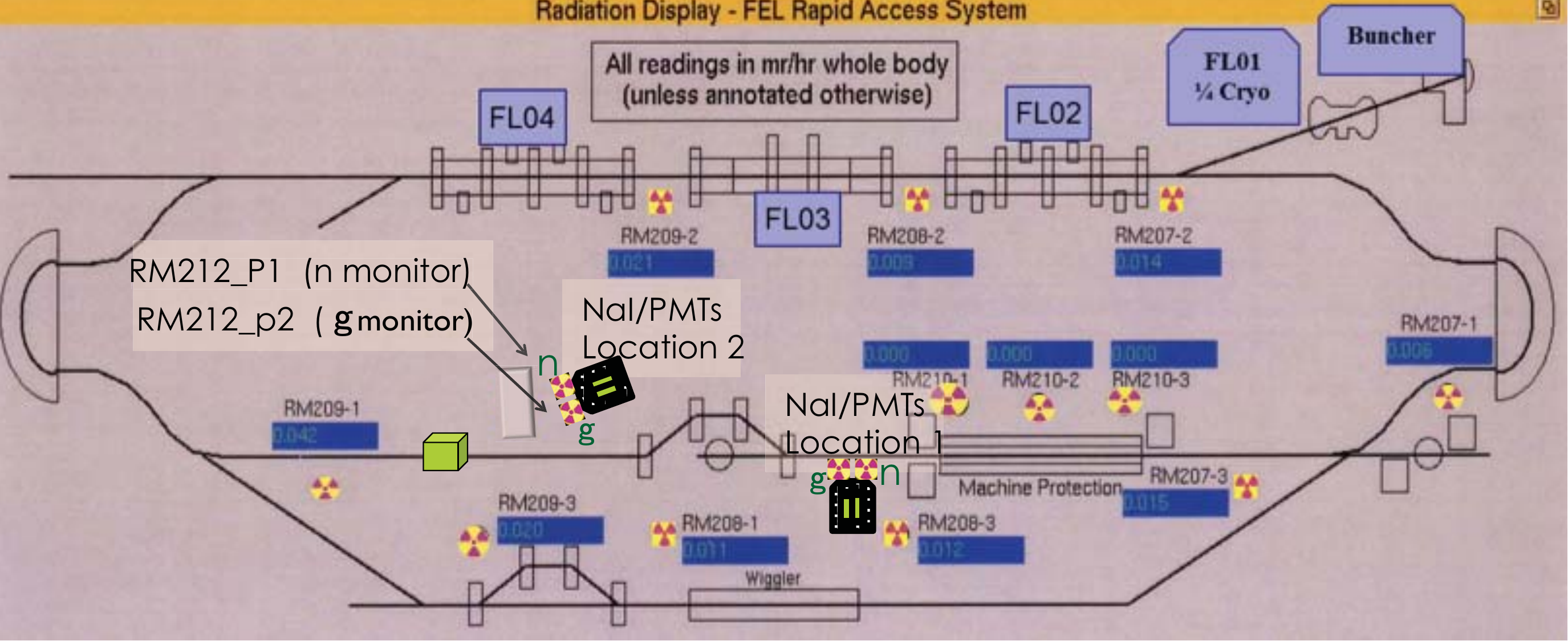}
\caption{Background radiation monitor layout}
\label{DL_transp3}
\end{figure}

\subsection{Beam setup}

The beam requirements for the transmission test were maximum average beam current with small momentum spread and r.m.s. beam radius $\ll1$ mm as well as minimal beam halo outside a 1 mm radius at the test aperture.
 
Details of the accelerator and beamline configuration for this test are discussed in ref. [3]. Small momentum spread was provided by "cross-phased" linac operation with the beam accelerated on the rising part of the RF wave in the first and third (low-gradient) accelerator module and on the falling part of the RF wave in the second (high-gradient) module. The phase-energy correlations so induced then cancel one another resulting in a small relative momentum spread of order 0.2\% f.w.h.m.

\subsubsection{Test Region Beam Optics}

Minimal size of the core beam at the aperture was achieved by two alternate-gradient quadrupole triplets up- and down-stream of the aperture, producing a "mini-beta" region of $\beta\approx0.2$ m and an r.m.s. beam radius of $\approx100$ $\mu m$ at the aperture. Additional quadrupoles and a full complement of beam monitors near the test region allowed beam phase advance adjustment, beam matching, and halo control without excessive betatron mismatch.

\subsubsection{Halo Management}

The moderate bunch charge of 60 pC minimized emittance and halo at the source. The small momentum spread alleviated the impact of dispersion errors, suppressed momentum tails, and mitigated effects of increased chromaticity of the "mini-beta" section. The longitudinal matching process (cross-phasing) allowed for a long bunch, reducing resistive-wall (wake field) effects in the aperture.

\subsubsection{Beam Tune}

Starting with a low-power beam, a longitudinal match, lattice dispersion, and betatron match were established and subsequently repeated  after each beam power increase. The "mini-beta" section was tuned by first centering and logging the beam orbit through the 6 mm and the 2 mm apertures. After inserting the beam viewers, the beam spot at the aperture position was then minimized (see Fig.4)

\begin{figure}[htbp]
\centering\includegraphics[width=0.23\textwidth,height=0.15\textheight]{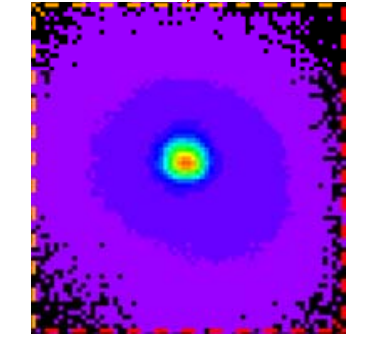}
\caption{Optimal beam profile at the aperture position, frame size = 1 mm x 1 mm}
\label{DL_transp4}
\end{figure}

The beam down-stream of the test region was then retuned for low loss and zero beam break-up (BBU) resulting from unstable beam oscillations.

Subsequently, several combinations of bunch charges and repetition rates were tested for minimal background and aperture block heating in transmission through the 4 and 2 mm apertures. Finally with fixed 60 pC bunch charge, beam transmission through the 2 mm aperture was optimized for increasing steps of beam power, fine-tuning the the beam after each step, until it reached its full power of about 450 kW (4.5 mA, 100 MeV).

\subsubsection{Transmission Runs}

Halo losses monitored by ion chambers and PMT's at the linac were kept minimal by tuning. Fine adjustment of beam steering and focussing near the aperture region kept the aperture block temperature rise and neutron and photon backgrounds minimal. Although a round beam spot of $50$ $\mu m$ radius was achieved, $100$ $\mu m$ spot radii at the aperture were typical.

A novel feature was BBU caused by small energy shifts. Because of the small momentum spread of 0.2\%, the usually observed Landau damping of BBU by tune spread from much larger momentum spreads was absent. Small energy shifts coupled to the chromaticity shifted the vertical phase advance which led to the onset of BBU. Stability was maintained by monitoring and minimizing the vertical beam size (see Fig. 5).

\begin{figure}[htbp]
\centering\includegraphics[width=0.7\textwidth,height=0.13\textheight]{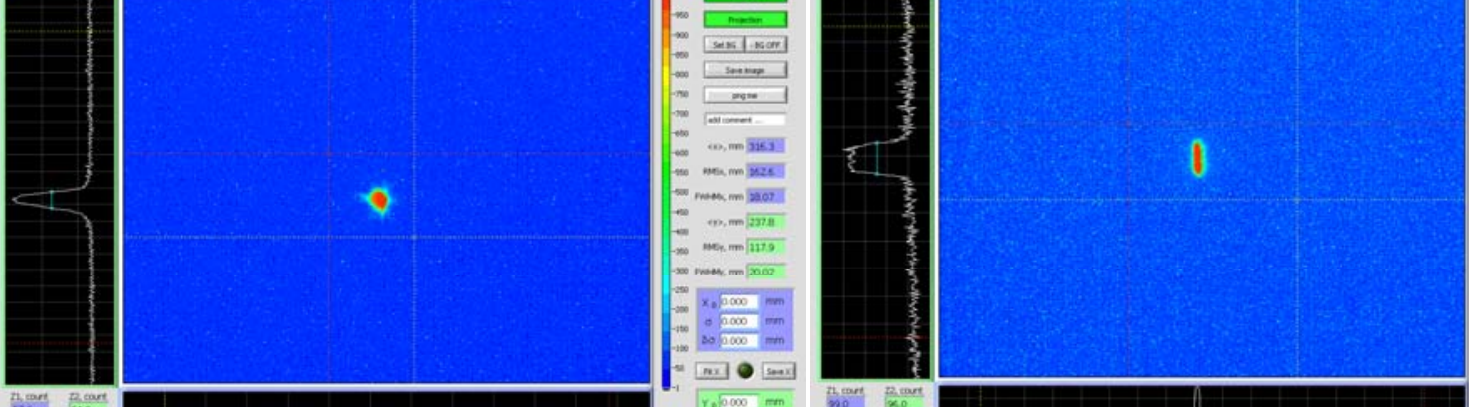}
\caption{Beam sizes at the synchrotron light monitor. Left: stable beam; right: beam near instability threshold}
\label{DL_transp5}
\end{figure}

In the final 7-hour transmission run through the 2 mm aperture, the machine instabilities (particularly BBU) were controlled by stabilizing the energy at injection and in the recirculator.

\section{Test Results}

After optimzing the beam transmission through the apertures, four runs were recorded: Nr. 1 and 2 of 22 minutes and 30 minutes through the 6 mm and 4 mm apertures and Nr. 3 and 4 of 124 minutes and 413 minutes through the 2 mm aperture. The time logs are shown in Figs. 6-9; Fig. 10 shows the log of a cooling period of the aperture block after the beam had been turned off (magenta traces indicate the aperture block temperature). 

\begin{figure}[htbp]
\centering\includegraphics[width=1.0\textwidth,height=0.4\textheight]{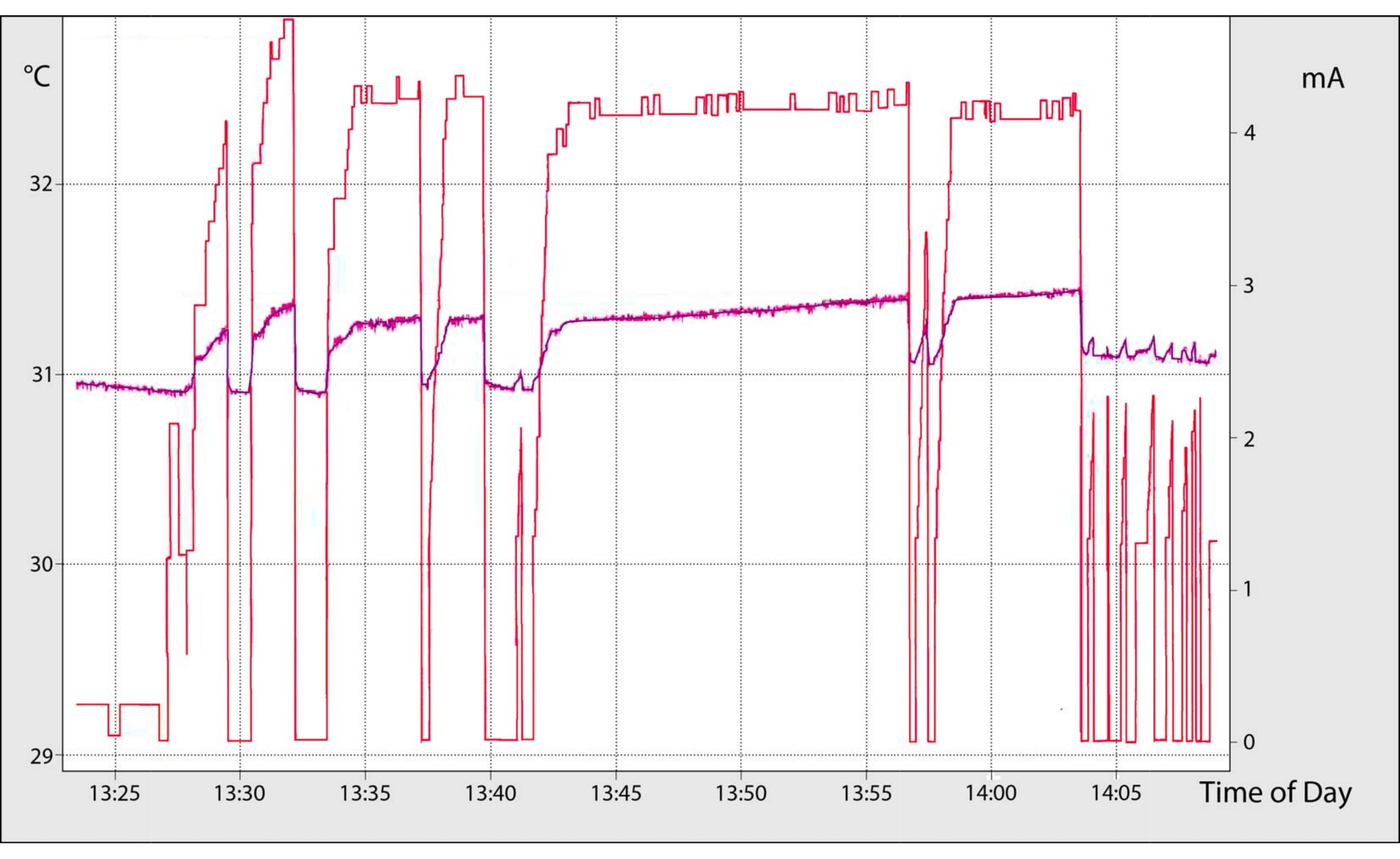}
\caption{Run 1 (6-mm aperture): block temperature (magenta) and beam current (red)}
\label{DL_transp6}
\end{figure}

\begin{figure}[htbp]
\centering\includegraphics[width=0.93\textwidth,height=0.40\textheight]{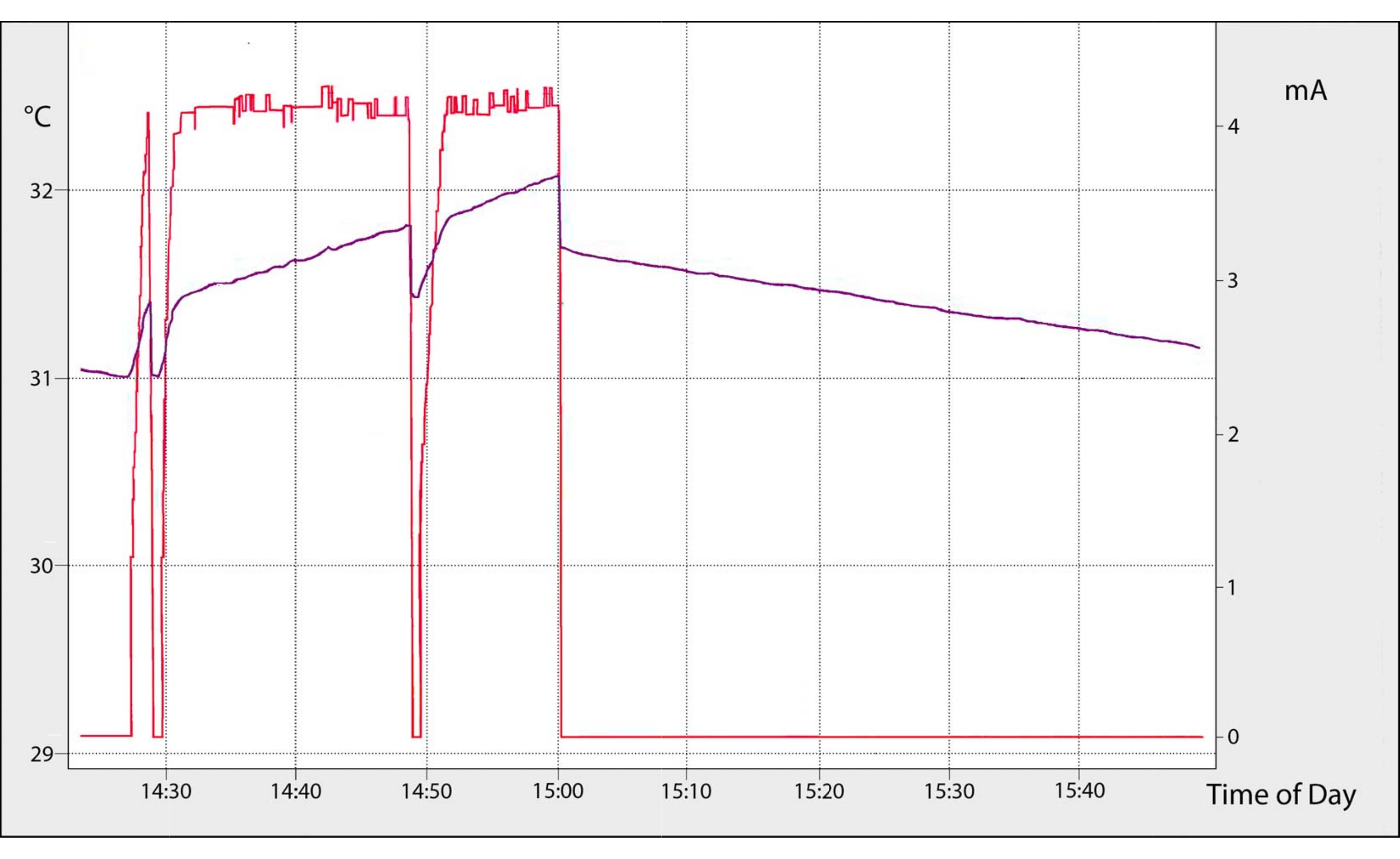}
\caption{Run 2 (4-mm aperture): block temperature (magenta) and beam current (red)} 
\label{DL_transp7}
\end{figure}

\begin{figure}[htbp]
\centering\includegraphics[width=0.9\textwidth,height=0.40\textheight]{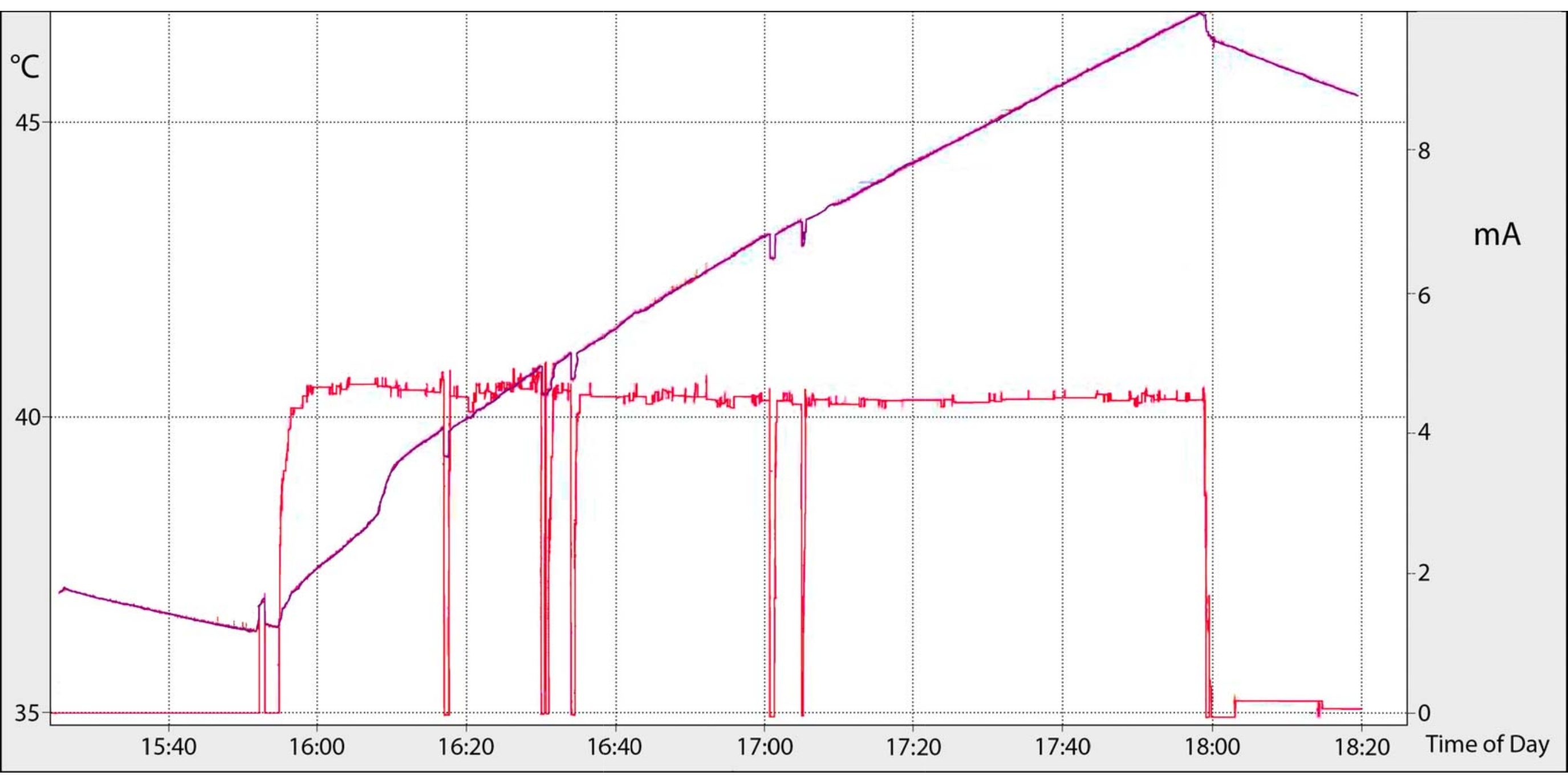}
\caption{Run 3 (2-mm aperture): block temperature (magenta) and beam current (red)}
\label{DL_transp8}
\end{figure}

\begin{figure}[htbp]
\centering\includegraphics[width=0.9\textwidth,height=0.45\textheight]{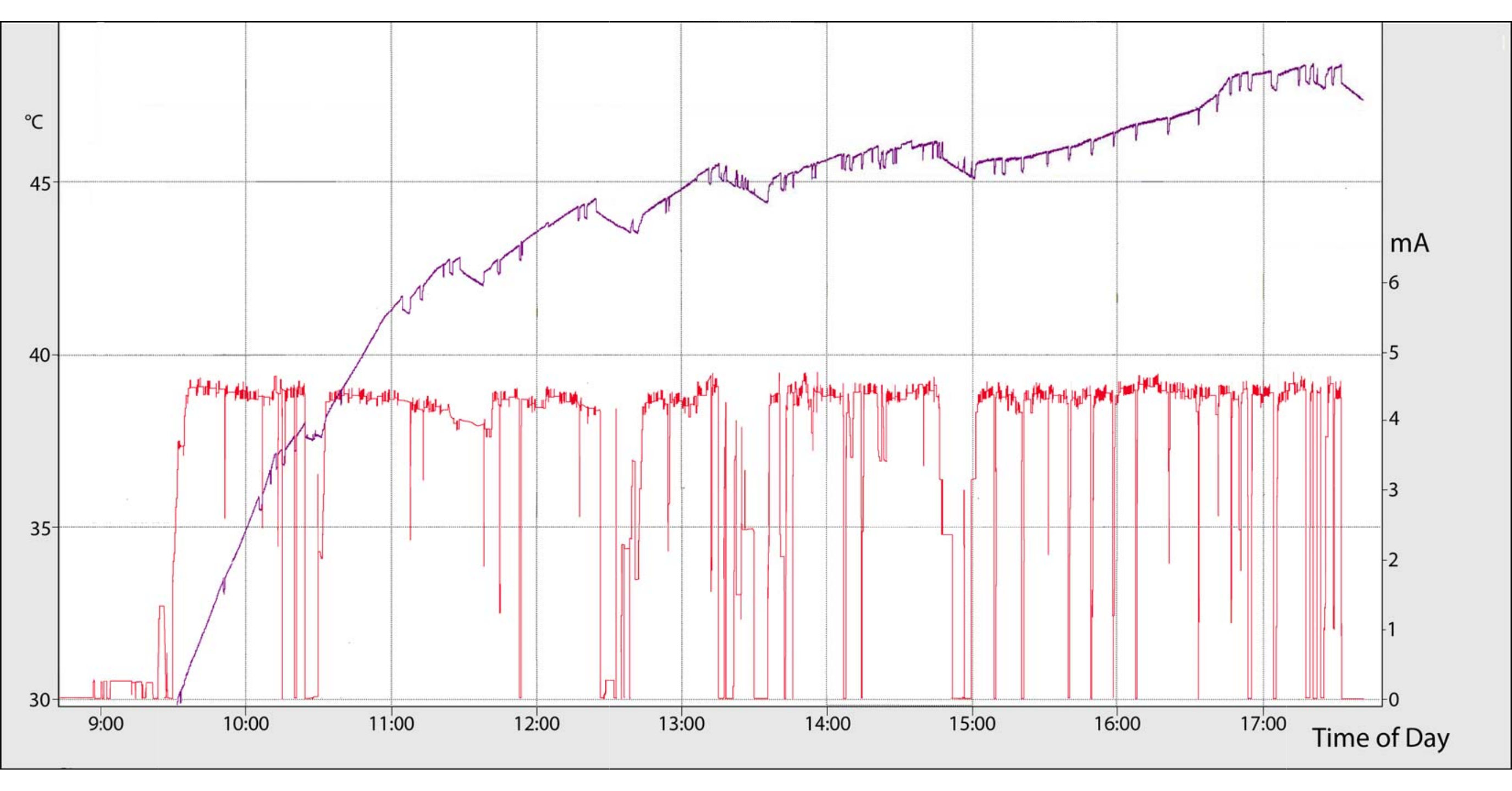}
\caption{Run 4 (2-mm aperture): block temperature (magenta) and beam current (red)}
\label{DL_transp9}
\end{figure}

\begin{figure}[htbp]
\centering\includegraphics[width=0.9\textwidth,height=0.4\textheight]{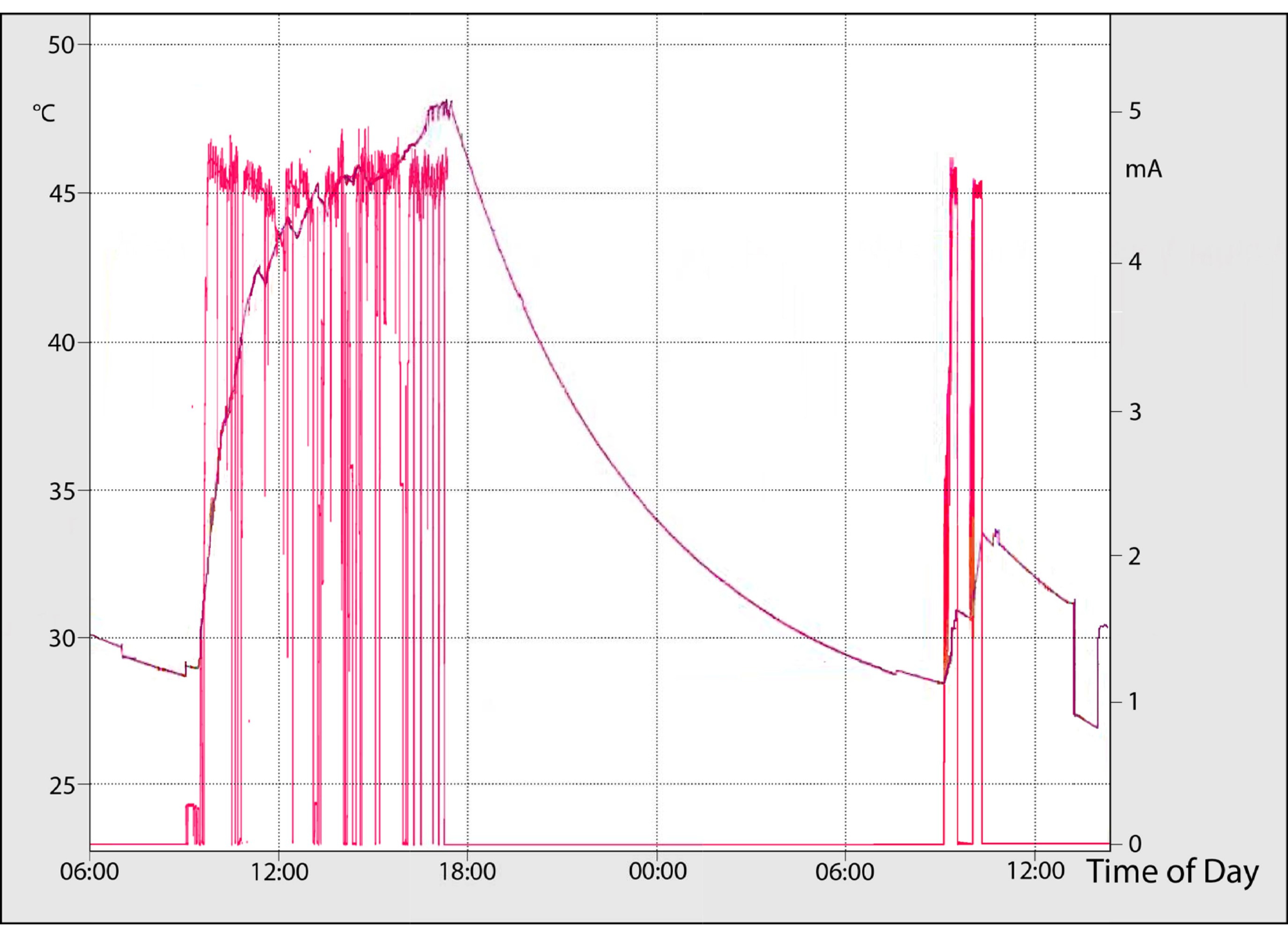}
\caption{Cooling runs without beam: block temperature (magenta), beam current (red)}
\label{DL_transp10}
\end{figure}

\newpage

\subsection{Beam Power Loss}

The power $P_B$ deposited by the beam in the aperture block is

\begin{equation}
P_B=c_pm(dT/dt)_{block}+P_C
\end{equation}

where $c_p$ and $m$ are the heat capacity and mass of the block $(c_pm=917Joule/^oC)$ and $(dT/dt)_{block}$ is the temperature change of the block. $P_C$ is the power lost from the block through heat conduction and radiation to the beam pipe.

From the cooling runs (Fig.10), we obtained a fit to the temperature $T(t)$ of the cooling block (beam turned off at t = 0):

\begin{equation}
T(t)=T_0+[T(0)-T_0]e^{-t/\tau}
\end{equation}

with $T_0=27.2$ $^oC$ and $\tau=357$ minutes. The cooling power $P_C$ is therefore

\begin{equation}
P_C=-c_pm(dT/dt)=c_pm(T-T_0)/\tau=0.0428(W/^oC)\cdot(T-T_0)
\end{equation}

and thus

\begin{equation}
P_B=c_pm[(dT/dt)_{block}+(T-T_0)/\tau]
\end{equation}

The integrated beam energy deposited in the block during a run from time $t_1$ to $t_2$ is

\begin{eqnarray}
E_B=\int_{t_1}^{t_2}dt\cdot P_B &=& c_pm[\Delta T+(T_{ave}-T_0)\Delta t/\tau]\\ &=& 917(Joule/^oC)[\Delta T+(T_{ave}-27.2^oC)\cdot0.0028\Delta t/min]
\end{eqnarray}

where $\Delta T$ is the temperature rise $T(t_2)-T(t_1)$ during the run time $\Delta t=t_2-t_1$ and $T_{ave}$ is the average temperature during the run. The temperature rise $\Delta T$ and the average temperature $T_{ave}$, average deposited power $P_B$, and beam power $P_b$ as well as the total charge and average beam current for each run are summarized in Table 1.

\begin{table}[htbp]
\centering
\begin{tabular}{|c|c|c|c|c|c|c|c|c|}
\hline\hline
Run & apert. & duration  & $\Delta T(^oC)$ & $T_{ave}(^oC)$ & $P_B(W)$ & $P_b(MW)$ & charge(C) & $I_{ave}(mA)$ \\
\hline\hline
1 & 6 mm & 22 min & 0.21  & 31.4  & 0.32  & 0.384  & 5.06 C & 3.84 \\
2 & 4 mm & 30 min & 0.65  & 31.6  & 0.52  & 0.393  & 7.08 C & 3.93 \\
3 & 2 mm & 124 min & 10.5  & 42.6  & 1.95  & 0.425  & 31.6 C & 4.25 \\
4 & 2 mm & 413 min & 9.1  & 44.8  & 1.09  & 0.422  & 121 C & 4.22 \\
\hline\hline
\end{tabular}
\caption{Transmission Results}
\label{Target_comp}
\end{table}

The power of the beam halo intercepted by the aperture block is only partly deposited in the block. A substantial part of the electromagnetic shower generated by the intecepted electrons is escaping through the back and the sides of the block. Modeling by the FLUKA code of a simplified aperture block (see Fig.15) showed that about 50\% of the energy of electrons entering the block near the 2 mm aperture is deposited in the block.

\subsection{Neutron Flux}

\subsubsection{Measured Neutron Fluxes}

The neutron fluxes from the aperture block were measured by a Canberra NP100B neutron rem-counter labeled rad212-p1 whose response function $c_n$ is shown in Fig. 17. It was positioned 1.9 m downstream of the aperture block and $24^o$ to the left of the beam axis. The fluxes for runs 1 to 4 are shown in Figs. 11-13.

\begin{figure}[htbp]
\centering\includegraphics[width=0.9\textwidth,height=0.22\textheight]{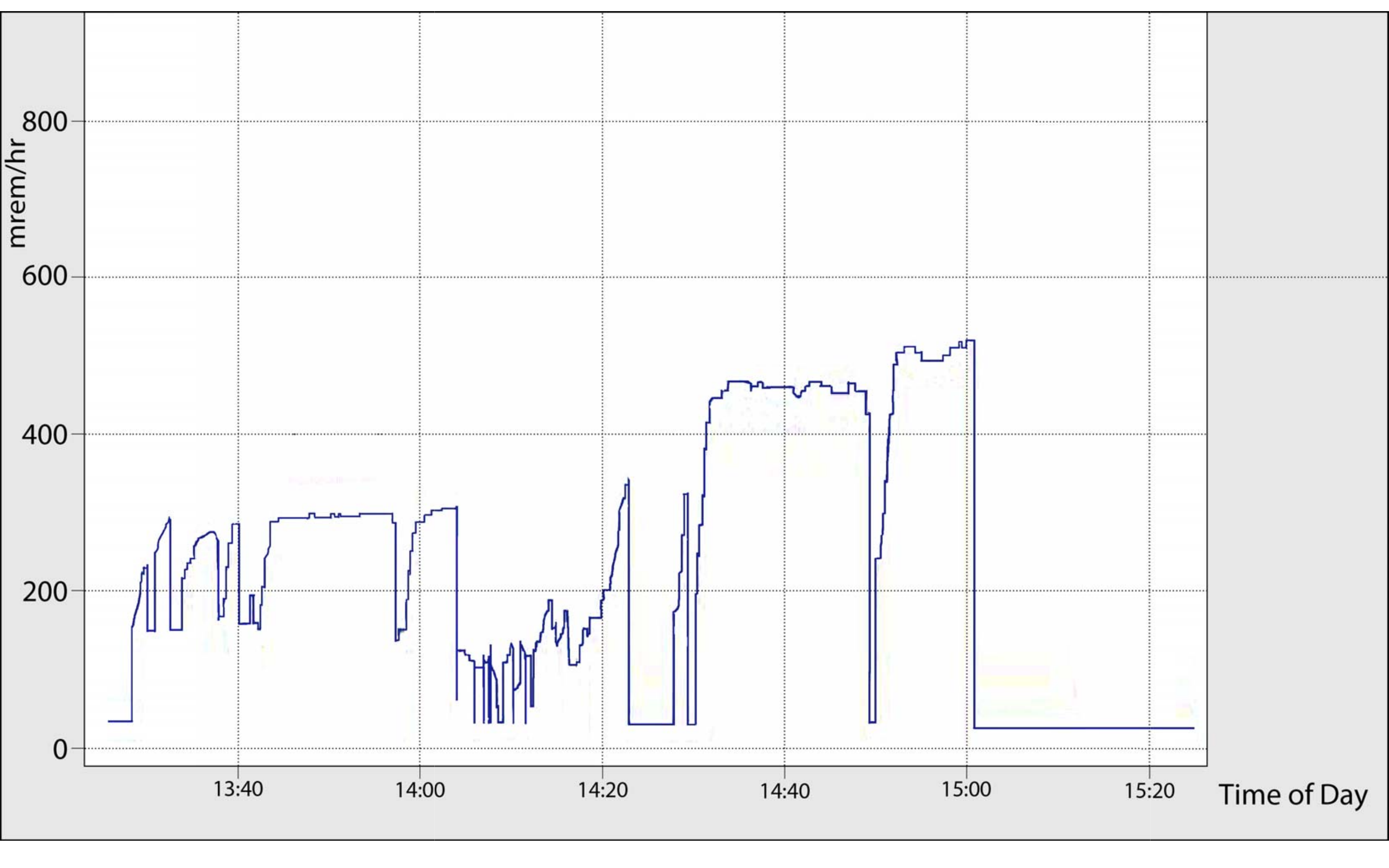}
\caption{Neutron flux for runs 1 and 2}
\label{DL_transp11}
\end{figure}

 \begin{figure}[htbp]
\centering\includegraphics[width=0.9\textwidth,height=0.22\textheight]{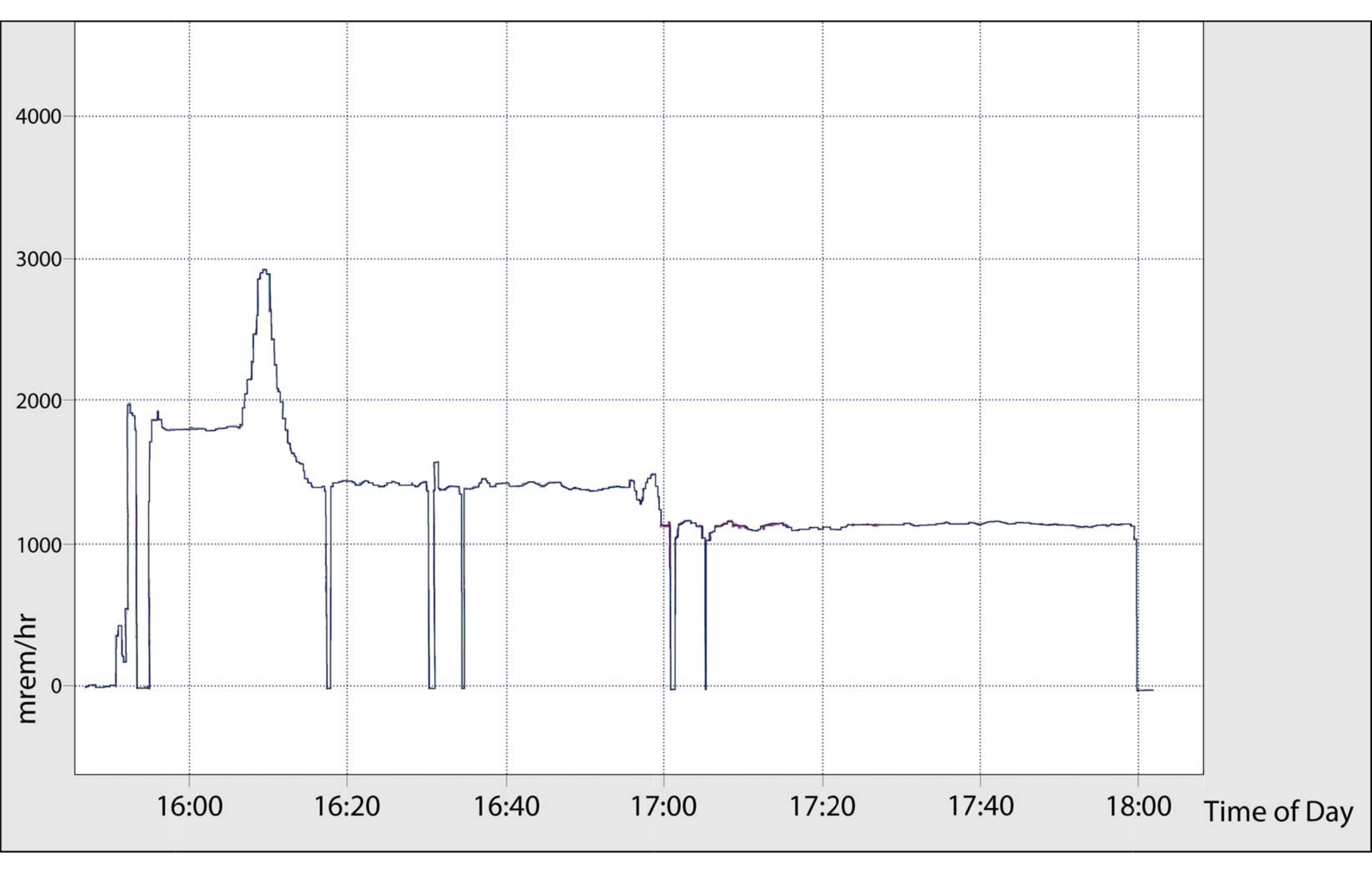}
\caption{Neutron flux for run 3}
\label{DL_transp12}
\end{figure}

\begin{figure}[htbp]
\centering\includegraphics[width=0.9\textwidth,height=0.22\textheight]{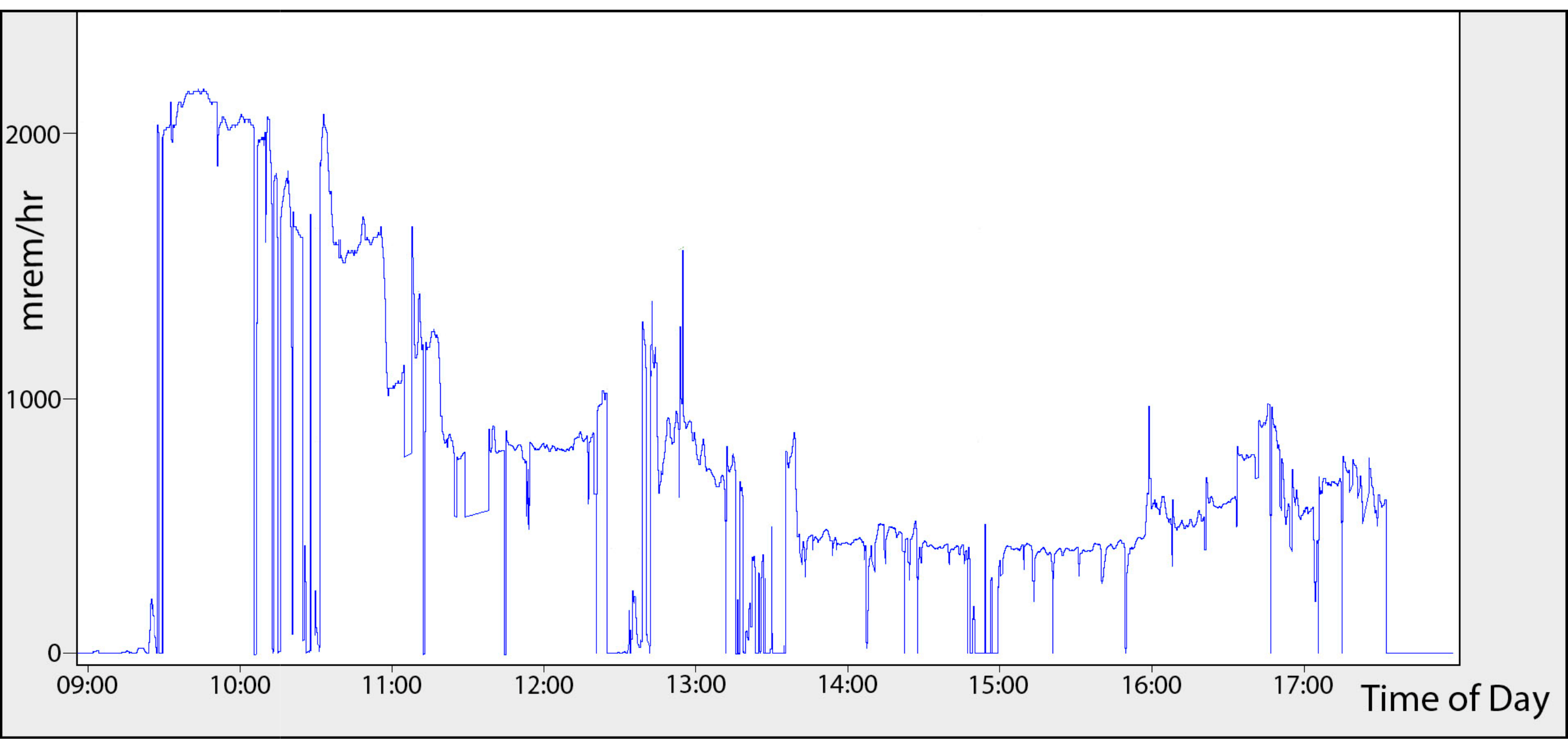}
\caption{Neutron flux for run 4}
\label{DL_transp13}
\end{figure}

\newpage
In order to relate the measured neutron fluxes to the power $P_B$ deposited in the block for a range of beam halo conditions, a selection of short sections of runs 3 and 4 where the neutron flux was reasonably stable were evaluated individually. The neutron dose rates $R_n$ plotted versus the block power $P_B$ are shown in Fig. 14.

\begin{figure}[htbp]
\centering\includegraphics[width=1.0\textwidth,height=0.45\textheight]{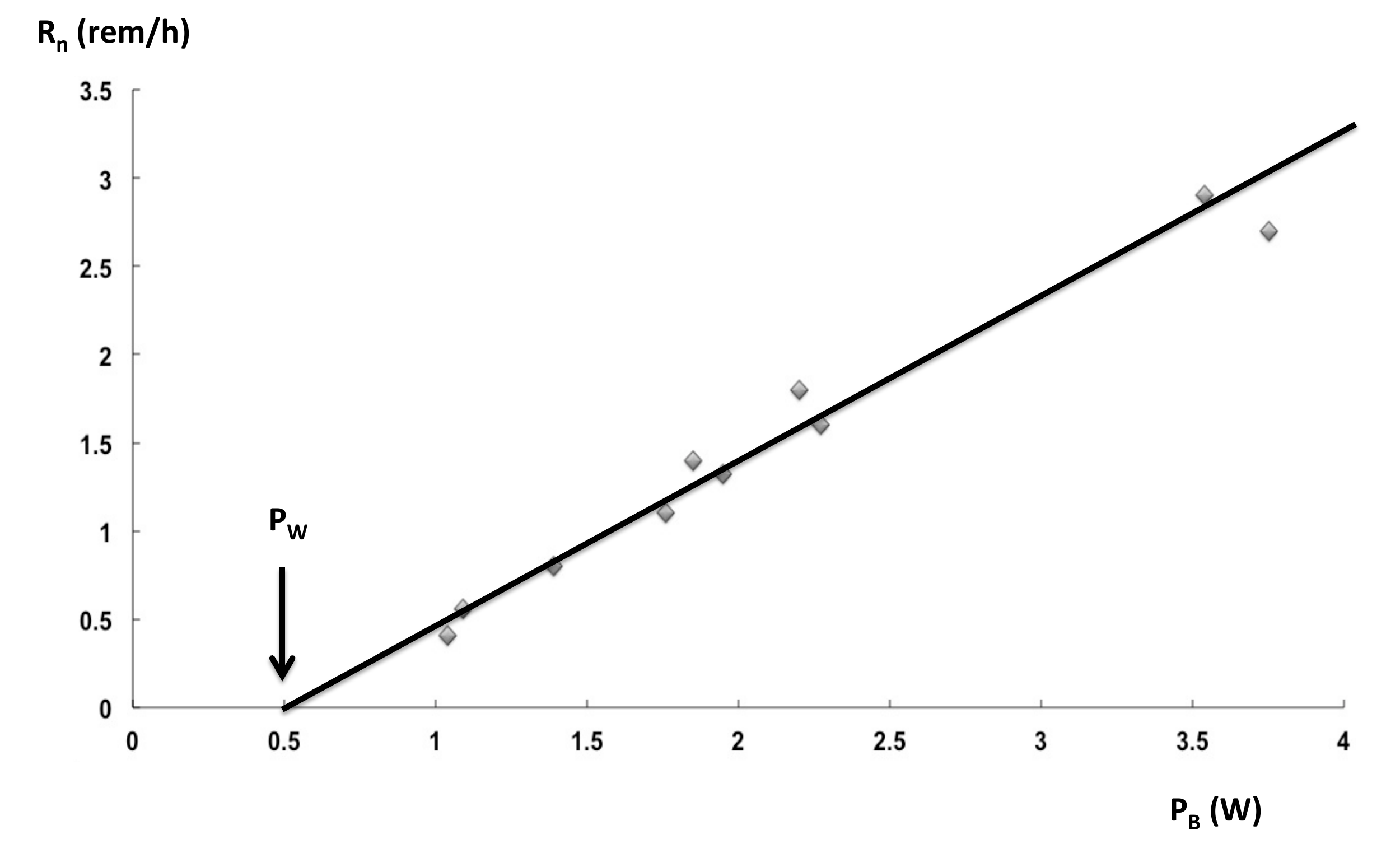}
\caption{Neutron dose rates vs. shower the block power $P_B$}
\label{DL_transp14}
\end{figure}

From the plot in Fig. 14, a linear relation was deduced of the form

\begin{equation}
R_n=0.9[rem/(Wh)]\cdot(P_B-P_W)
\end{equation}

where $P_W\approx0.5$ W was interpreted as the power deposited by the wake fields of the beam. Since the r.m.s. width of the beam at the aperture was less than about 0.1 mm or ten times smaller than the 2 mm aperture, it is reasonable to assume that the wake fields are largely governed by the bunch charge and time structure of the beam which were kept fixed throughout all four transmission runs.

Since FLUKA simulations have shown that the power $P_B-P_W$ deposited by the beam halo in the aperture block is only about 50\% of the total power $P_H$ of the intercepted beam halo, we have

\begin{equation}
P_H\approx2.0\cdot(P_B-P_W)
\end{equation}

and eqn. (7) becomes 

\begin{equation}
R_n\approx 0.45[rem/(Wh)]\cdot P_H
\end{equation}

The relevant results for the individual sections 3.1 to 3.5 of run 3 and 4.1 to 4.3 of run 4, as well as their averages labeled 3.0 and 4.0 for each run, are shown in Table 2.

\begin{table}[htbp]
\centering
\begin{tabular}{|c|c|c|c|c|c|c|c|c|}
\hline\hline
Run & start & stop  & $\Delta T(^oC)$ & $T_{ave}(^oC)$ & $P_B(W)$ & $R_n(rem/h)$ & $P_H(W)$ & $I_{ave}(mA)$ \\
\hline\hline
3.0 & 15:55 & 17:59 & 10.5   & 42.6  & 1.92  & 1.32  & 2.9  & 4.25  \\
3.1 & 14:51 & 14:54.7 & 0.785   & 34.1  & 3.54  & 2.8  & 6.1 & 4.25  \\
3.2 & 14:55 & 15:05 & 2.20   & 35.7  & 3.72  & 2.7  & 6.5  & 4.25  \\
3.3 & 15:56 & 16:08 & 1.30   & 37.7  & 2.10  & 1.8  & 3.2  & 4.3  \\
3.4 & 16:35 & 17:00 & 1.98   & 42.2  & 1.85  & 1.4  & 2.7  & 4.3  \\
3.5 & 17:04 & 17:59 & 3.58   & 45.2  & 1.76  & 1.1  & 2.5  & 4.2  \\
\hline
4.0 & 10:39 & 17:32 & 9.1   & 44.8  & 1.06  & 0.58  & 1.2  & 4.23  \\
4.1 & 10:39 & 10:59 & 2.25   & 40.0  & 2.27  & 1.6  & 3.6  & 4.3  \\
4.2 & 11:38 & 12:23 & 2.04   & 43.5  & 1.39  & 0.8  & 1.8  & 4.3  \\
4.3 & 15:00 & 16:00 & 0.93   & 46.0  & 1.04  & 0.4  & 1.1  & 4.3  \\
\hline\hline
\end{tabular}
\caption{Block power and total shower power vs. neutron flux}
\label{Target_comp}
\end{table}

Applying the same relation $R_n(P_B,P_W)$ of eqn. (7) to runs 1 and 2 for 6 mm and 4 mm apertures, the resulting wake field powers $P_W$  deposited are 0.1 W for the 4 mm and 0.075 W for the 6 mm aperture.

\subsubsection{Neutron Flux Modeling}

The neutrons are generated mainly by  Giant Resonance interactions of the primary electron and the associated electromagnetic shower with the target nuclei. However, since the aperture block shown in a simplified form in Fig. 15 is too short and too narrow to absorb the entire electromagnetic shower produced by the intercepted electrons, the remaining shower escapes through the sides and the downstream face of the block. It propagates down the beam pipe and is eventually absorbed in the pipe and surrounding beam line components, producing additional neutrons downstream of the aperture block and closer to the neutron detector (see Fig. 16).

\begin{figure}[htbp]
\centering\includegraphics[width=0.45\textwidth,height=0.18\textheight]{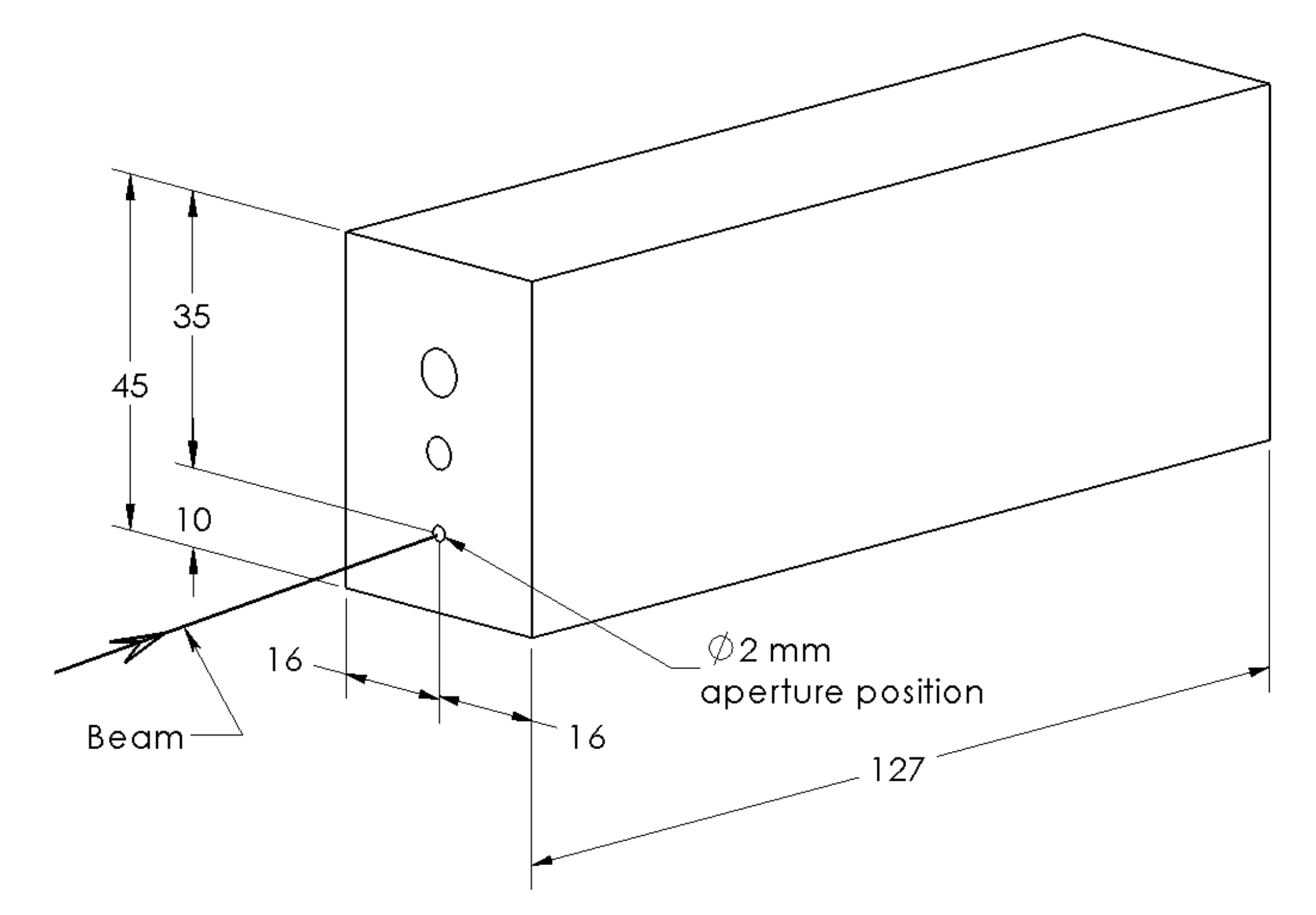}
\caption{Simplified aperture block}
\label{DL_transp15}
\end{figure}

\begin{figure}[htbp]
\centering\includegraphics[width=0.5\textwidth,height=0.19\textheight]{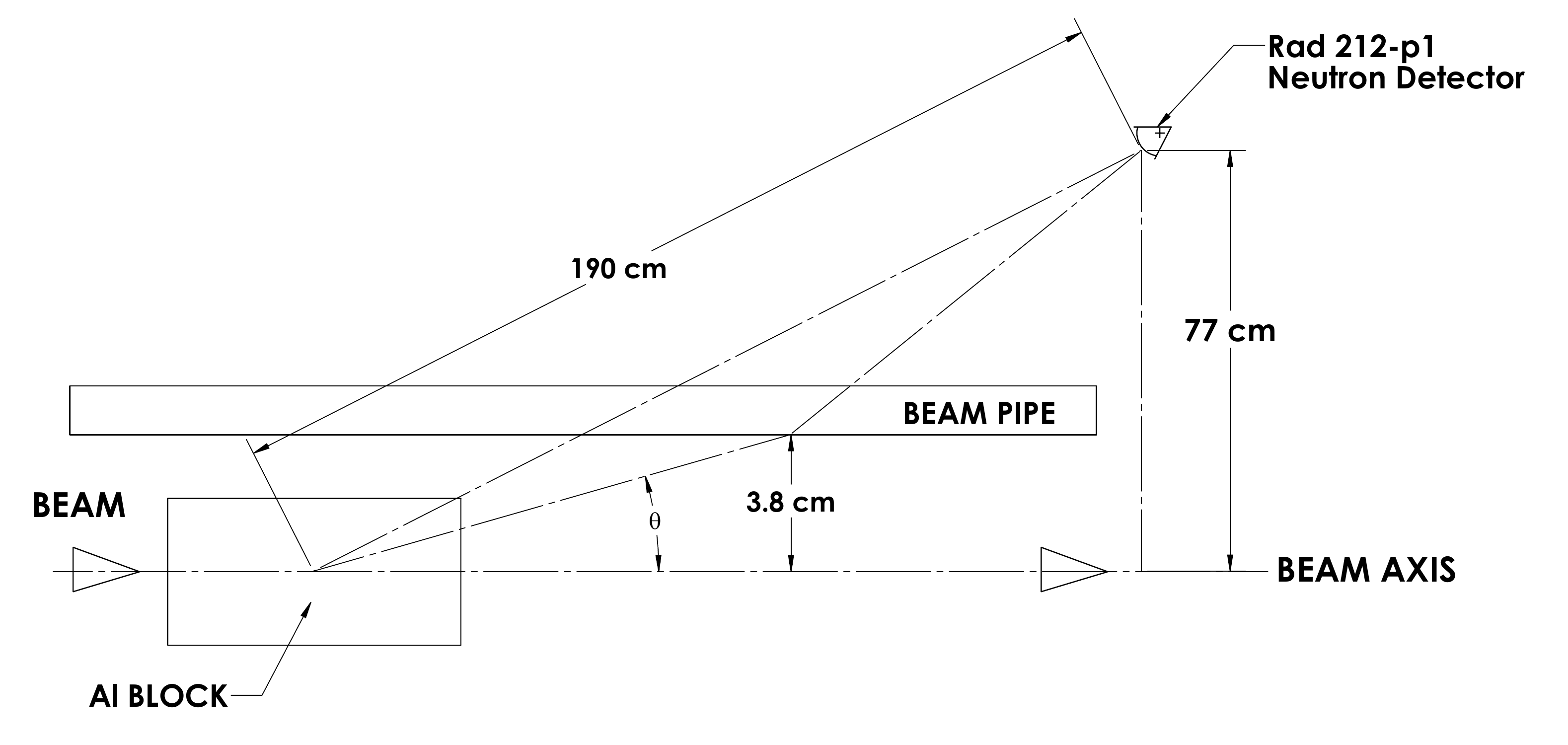}
\caption{Neutron generation geometry}
\label{DL_transp16}
\end{figure}

\newpage

For the simplified aperture block of Fig. 15 placed inside a thick steel pipe of 7.6 cm ID and 12 cm OD and with a 100 MeV electron "pencil" beam entering the block at the position of the 2-mm aperture, the resulting neutron flux at the rad212-p1 detector was modelled using the MCNP code. The resulting neutron energy spectrum for $10^7$ incident electrons is shown in Fig. 17.

\begin{figure}[htbp]
\centering\includegraphics[width=1.0\textwidth,height=0.35\textheight]{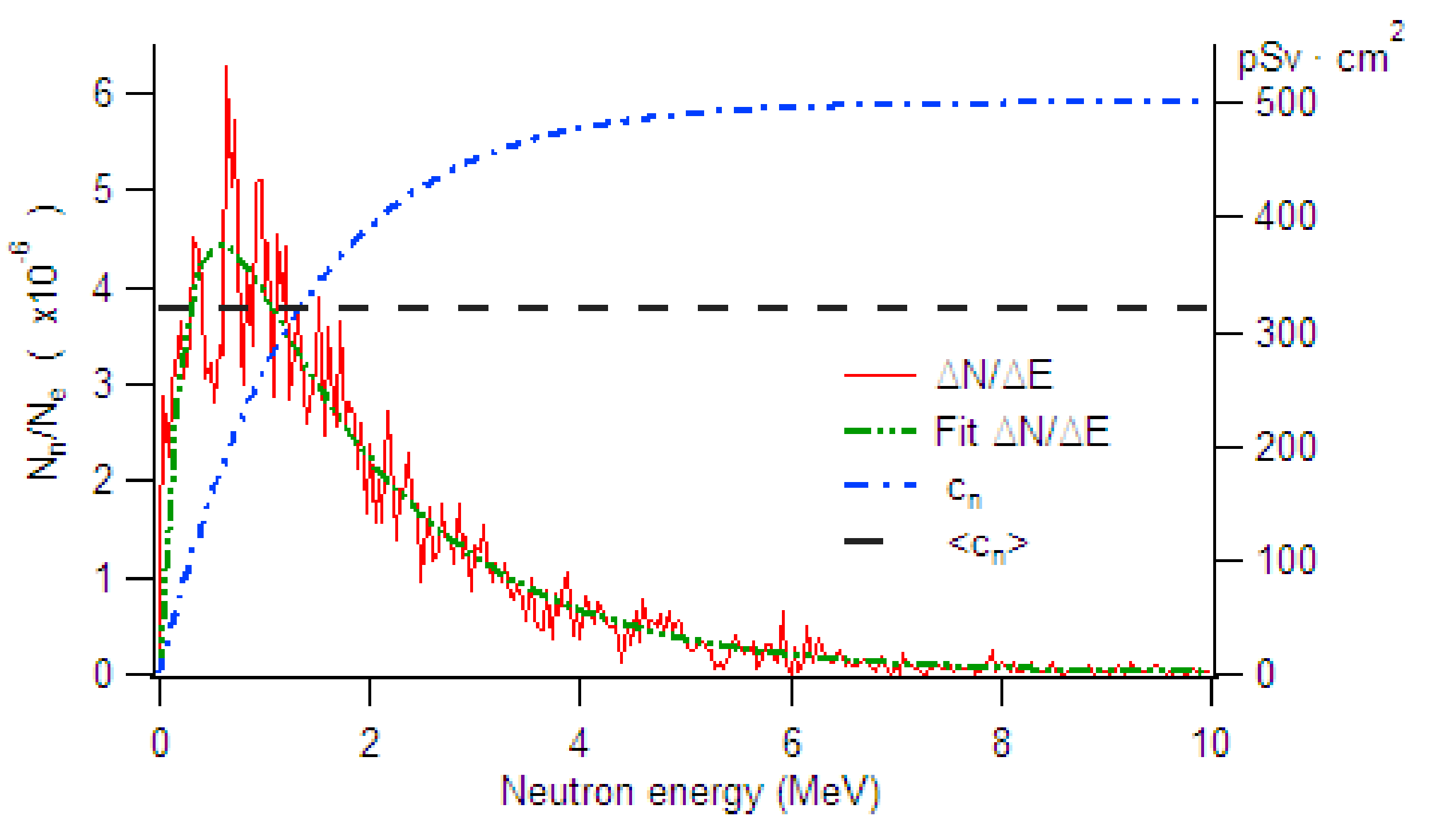}
\caption{Neutrons per incident electron and per MeV at the rad212-p1 detector (MCNP simulation) and the effective dose conversion factor $c_n$ of the neutron detector from ref. [4]}
\label{DL_transp17}
\end{figure}

The integral flux at the detector was $1.05\pm0.01\cdot10^{-5}$ neutrons per electron into a sphere of 8 cm radius, amounting to an integral flux density of $5.2\cdot10^{-8}$ neutrons/$cm^2$ or

\begin{equation}
\frac{dN_n}{dA\cdot dt}=3300\frac{neutrons}{cm^2Ws}\cdot P_H
\end{equation}

The fit to the neutron energy spectrum in Figure 17 has the form

\begin{equation}
d(N_n/N_e)/dE=[0.005+0.91e^{-E/1.45MeV}-0.8e^{-E/0.4MeV}]\cdot10^{-5}/MeV
\end{equation}

In order to compare the measured flux with these model predictions, the neutron energy spectrum has to be folded with the response function, i.e. the effective dose conversion factor $c_n$, of the neutron detector to obtain the expected neutron dose rate

\begin{equation}
R_n=\int\frac{dN_n}{dA\cdot dt\cdot dE}\cdot c_n(E)dE
\end{equation}

The function $c_n(E)$ shown in Fig. 17 was taken from ref. [4] and parametrized as

\begin{equation}
c_n=500(pSv\cdot cm^2)\cdot(1-e^{-E/1.3MeV})
\end{equation}

The resulting effective value of $c_{n,aver}=320$ $pSv\cdot cm^2$ yielded  the relation

\begin{equation}
R_n\approx0.38[rem/(Wh)]\cdot P_H
\end{equation}

which is only about 15\%  below the measured value of eqn. (9).
 
\section{Conclusion}

Table 3 shows a summary of average results for the four transmission runs.

\begin{table}[htbp]
\centering
\begin{tabular}{|c|c|c|c|c|c|c|c|c|c|}
\hline\hline
Run & start & stop  & $\Delta T(^oC)$ & $T_{ave}(^oC)$ & $P_B(W)$ & $P_W(W)$ & $R_n(rem/h)$ & $I_{ave}(mA)$ & beam loss \\
\hline\hline
1 & 13:42 & 14:03 & 0.212 & 31.35 & 0.33 & 0.075 & 0.24 & 3.84 & 1.3 ppm \\
2 & 14:32 & 15:01 & 0.65 & 31.55 & 0.52 & 0.10 & 0.43 & 3.93 & 2.1 ppm \\
3.0 & 15:55 & 17:59 & 10.5 & 42.6 & 1.95 & 0.50 & 1.32 & 4.25 & 6.8 ppm \\
4.0 & 10:39 & 17:32 & 9.1 & 43.8 & 1.09 & 0.50 & 0.58 & 4.23 & 2.8 ppm \\
\hline\hline
\end{tabular}
\caption{Wake field power and beam transmission losses}
\label{Target_comp}
\end{table}

The most significant result of these transmission tests was that we succeeded in running a 100 MeV electron beam of 0.43 MW average power for 7 hours through a 2 mm diameter aperture of 127 mm length with an average beam loss of about 3 ppm, with an estimated uncertainty of $\pm20\%$.

\section{Acknowledgements}

We are grateful for the design and construction of the target assembly by the MIT-Bates Research and Engineering Center and for the accelerator preparation and skilful beam delivery by the FEL crew of the Thomas Jefferson National Accelerator Facility. The research is supported by the United States Department of Energy Office of Science.

Notice: Authored by Jefferson Science Associates, LLC under U.S. DoE Contract No. DE-AC05-060R23177. The U.S. Government retains a non-exclusive, paid-up, irrevocable, world-wide licence to publish or reproduce this manuscript for U.S. Government purposes. This work supported by the Commonwealth of Virginia, by DoE under contract DE-AC05-060R23177, and by MIT Nuclear under DoE contract DE-FG02-94ER40818.

\section{References}

[1] R. Alarcon et al., "Transmission of Megawatt Relativistic Electron Beams Through Millimeter Apertures", arXiv: physics.acc-ph/1305.0199, May 1, 2013; submitted to Physical Review Letters.\\\ [2] R. Alarcon et al., "Measured Radiation and Background Levels During Transmission od Megawatt Eletron Beams Through Millimeter Apertures", arXiv:\\\
[3] D. Douglas et al., "IR FEL Driver ERL Configuration for the DarkLight Aperture Test", JLAB-TN-13-009, 11 December 2012, and  D. Douglas et al.,"Accelerator Operations for DarkLight Aperture Test", JLAB-TN-13-020, 16 April 2013.\\\  [4] M. Pellicioni, Radiation Protection Dosimetry 88, 279 (2000).

\end{document}